\documentclass[12pt]{article}
\usepackage{authblk}
\usepackage{bm}
\usepackage[english]{babel}
\usepackage{lineno}

\begin{document}
	
	\noindent
	{{\bf \Large {Einstein-Grossmann geometry for frequency shifts of retrograde and direct orbits }}}
	
	\medskip \noindent {Igor {\'E}. Bulyzhenkov$^{1,2}$} 
	
	\medskip	\noindent 
		$^1$Space Research Institute RAS, 
		Moscow, Russia\\		
		$ ^2$Peoples' Friendship University of Russia, 
	Moscow, Russia
	\\	e-mail: ibphys@gmail.com

		\medskip
\noindent  {\bf Abstract.	}
The metric gyro-potential of rotating distributions creates centripetal forces that can override Newtonian attraction on the inner and near-zone orbits. Einstein\rq{s} geodesics in four metric potentials predict Zeeman-like shifts of Keplerian frequencies and measurable differences in the orbital speeds of direct and retrograde circulations.







	\medskip
\section {Einstein\rq{s} geodesic relations}

The difference in angular frequency of direct and retrograde orbits around rotating planets, stars and galaxies can challenge Newton\rq{s} incomplete potential in the low-speed region of its application.  Following the geometric insight of gravitational physics \cite {EinGro1913},
the metric four-velocity $cu_\mu \equiv cg_{\mu\nu} dx^\nu/ds \equiv cg_{\mu\nu}u^\nu $ of the probe mass $m_p$ (with $v^i =  dx^i/d\tau, v_i = g_{i\nu}dx^\nu/d\tau,$ $c^2 d\tau^2 = dx_\mu dx^\mu  + dl^2$) obeys both the geodesic relations $m_pc^2u^\nu \nabla_\nu u_\mu = 0 $  and the metric equality $m_pc^2u^\nu \nabla_\mu u_\nu = 0 $ from the defining relations $u^\nu u_\nu = 1$ and $ds^2 = g_{\mu\nu} dx^\mu dx^\nu = d x_\mu d x^\mu$.     
Einstein and Grossmann \cite {EinGro1914,Ein} relied on the symmetrical connections ${\Gamma^\lambda_{\mu\nu}} ={\Gamma^\lambda_{\nu\mu}} $ in the covariant derivative, $\nabla_\mu u_\nu \equiv \partial_\mu u_\nu - {\Gamma^\lambda_{\mu\nu}} u_\lambda $. Thus one can rewrite the celebrated geodesic relations for point masses without spin in metric fields 
by four partial derivatives $\partial_\mu \equiv \partial /\partial x^\mu$. This simplification for the antisymmetric tensor with covariant derivatives leads to the Lorentz (or electromagnetic) form of 3-accelerations in the 4-vector geodesic relations:
\begin {eqnarray}
\cases {
	0 = m_p c^2u^\nu (\nabla_\nu u_\mu - \nabla_\mu u_\nu) = m_p c^2u^\nu (\partial_\nu u_\mu - \partial_\mu u_\nu)\ $or $  
	\cr 0 =
	-m_p \gamma  v^j [c\partial_j (- \gamma {\sqrt {g_{oo}}}) - \partial_o \gamma (v_j - cg_j)]
	\Rightarrow - m_p \gamma g_o {\bm \beta} \cdot {\bm E} \
	$\ and $ \cr
	0\!=\! m_p u^o  [c^2\partial_i ( - {\gamma\sqrt {g_{oo}}})\! -\! c \partial_o \gamma ( v_i\! - \!c g_i)]\! 
	+\! m_p \gamma v^j
	[\partial_i \gamma ( v_j\! -\! c g_j)\! -\! \partial_j\gamma (v_i \!-\! c g_i) ] \cr  \Rightarrow m_p u^o [c^2 {\bm \partial} (- \gamma {\sqrt {g_{oo}}})\! -\! 
	c\partial_o ( \gamma{\bm v}\!-\!c \gamma{\bm g})  ] 
	\ +				   m_p\gamma {\bm v} \times curl [( {\bm v}\! - \!c {\bm g}) \gamma ]  \cr = m_p  (u^o g_o {\bm E} +  \gamma {\bm {\beta}} \times {\bm B} ).		
}
\end{eqnarray}
Here the 3-vector 
$\{curl \ {\bm f}\}^i \equiv e^{ijk} (\partial_j f_k - \partial_k f_j)/ 2 {\sqrt {|g_{ps}|} } $ is dual to the antisymmetric tensor $(\nabla_j f_k - \nabla_k f_j)$, 
$ \{{\bm B}\}^i =\{ c^2 curl (\gamma {\bm \beta} - \gamma {\bm g})\}^i $, $g_\nu \equiv g_{o\nu}/\sqrt {g_{oo}}$, 
$\{{\bm E}\}_i \equiv E_i = [c^2\partial_i ( - {\gamma\sqrt {g_{oo}}})\! -\! c \partial_o \gamma ( v_i\! - \!c g_i)]/g_o$, 
$u^i \equiv \gamma \beta^i, u_i = -\gamma(\beta_i-g_i)$, $v^j/c = \beta^j = dx^j/g_\nu dx^\nu,$  $v_i/c =\beta_i = (g_i g_j - g_{ij})dx^j/g_\nu dx^\nu $,  $ds^2 = g_{\mu\nu}dx^\mu dx^\nu = (cd\tau)^2 /\gamma^2 =  (g_\nu dx^\nu)^2(1-\beta_i\beta^i).$ 
For the metic-kinetic 3-fields ${E_i}$ and ${B^i}$  in the Einstein\rq{s} geodesic relations (1)  we assume that a quasi-isolated metric self-organisation is formed by a very heavy mass integral $M$, with $M \gg m_p$.  The inertial densities of the distributed mass-energy integral $Mc^2$ can rotate stationary around the axis z, while the probe mass $m_p$ can also circulate in the meridian mid-plane $z=0$ with $\gamma { v^i} = \{{\hat {\bm z}} \omega(r) \times {\bm r}\}^i =r\omega(r)$.

\section {Exact solutions for mid-plane orbits}

The cylindrical symmetry of the stationary galactic densities, for example, allows only two mid-disk components of the metric 4-potential 
$c^2g_\mu (x^o,r,\alpha, z=0)= c^2\{g_o(r);  0;  {\hat {\bm \alpha}}g(r); 0   \} $.  Such  a stable dynamical organization
maintains the radial kinetic-metric strength ${\bm E } \equiv {\hat {\bm r}}{E_r}(r, z=0) = - {\hat {\bm r}}[\partial_r (\gamma g_o )]c^2/ g_o\neq 0$ for  equatorial circulations of probe bodies with 
$m_p\partial_o ( \omega {\hat {\bm z}}\times {\bm r}  - {\hat {\bm \alpha}}\gamma g ) = 0 $ and $m_p {\gamma {\bm \beta} \times {\bm B}} \neq 0$.


Direct and retrograde orbits of probe masses in relativistic fields can be analytically described in exact mathematical terms of Einstein\rq{s} geodesic relations (1) for the simplest relativism of stationary circulations: 
\begin{eqnarray}
\cases { 
	0 = u^og_o	(E_r /c^2)\!+ \![ {\hat {\bm \alpha}} \gamma \beta \!\times\! curl ({\hat {\bm \alpha}} \gamma \beta - {\hat {\bm \alpha}}c\gamma g )  ]_r 
	\cr	= -{(1\!-\!g\beta)\gamma}\left ( \partial_r\!  \gamma \!+ \!\gamma \partial_r ln g_o \right)  
	+ 
	\frac {\gamma \beta }{r} \partial_r [r\gamma (\beta \!-  \!g) ]
	\cr =  g\beta\gamma \partial_r\!  \gamma \!-\! (1\!-\!g\beta)\gamma^2 \partial_rln g_o \!+\! 	\frac {\gamma^2 \beta(\beta\! -\!  g)  }{r} -\! \gamma \beta \partial_r (\gamma g) 
	\cr 
	= \frac {\gamma^2}{r} [ \beta^2 - \beta g (1+ r\partial_r ln [g/g_o])   - r\partial_r ln g_o ].
	\cr
} 
\end{eqnarray}
To simplify the summands on the right-hand side of stationary equation (2), we used  differential identities $\gamma\partial_r \gamma = \gamma \beta \partial_r (\gamma \beta)$  for the kinematic factor $\gamma = (1-\beta^2)^{-1/2}$.

The derived quadratic equation with respect the dimensional speed $\beta$ of geodesic circulations in stationary metric potentials $c^2g_o(r,z=0)$ and ${\hat {\bm \alpha}} c^2 g(r, z=0)$ gives as relativistic velocities for azimuthal co-rotation of probe masses and metric fields,   $v_{+} = c\beta_{+}> 0$, as well as the same orbit velocities of counter-rotation, $v_{-} = c\beta_{-}< 0$.  The Newton-Kepler regime for circular orbits, $v_{+} (r) \approx |v_{-} (r)|$, take place in the far zone of rotating densities, where   ${g^2[1+r(ln [g/g_o])^\prime_r]^2/4} \ll {r(ln g_o)^\prime_r} \approx v^2_+/c^2 \approx v^2_{-}/c^2 \ll 1$ 
and   
\begin {eqnarray}
\frac {v_{\pm}(r)}{c} \equiv\! 
\frac {g[1\!+\!r(ln [g/g_o])^\prime_r]}{2}\!
\pm\! \sqrt {  \frac {g^2[1\!+\!r(ln [g/g_o])^\prime_r]^2}{4}\! +\!r(ln g_o)^\prime_r} 
\cr	
\approx\cases{	
\sqrt { r(ln g_o)^\prime_r}\!+\! \frac {g[1+r(ln [g/g_o])^\prime_r]}{2}  >0,\! \cr
-\sqrt { r(ln g_o)^\prime_r}\!+\! \frac {g[1+r(ln [g/g_o])^\prime_r]}{2} < 0.
}
\end{eqnarray}

The Keplerian orbits around rotating masses $M$ can be modelled using Newton\rq{s} one-potential theory for the non-spinning probe mass $m_p$  ($g_o -1 \neq 0, g = 0$). There are no visual consequences from the vanishing gyro-potential  $g(R)$, which is much smaller than the Newtonian potential,  $1-g_o(R) = GM/c^2 R \ll 1 $, of distantly rotating masses $M$. In the so-called far zone, the incomplete Newtonian gravity  approximates Keplerian orbits  by equal direct and retrograde speeds:  $ c\sqrt {R\partial_R g_o (R)} \rightarrow \sqrt {GM/R}$. Zeeman-type perturbations of the far-zone orbits in (3) can only be studied through precise measurements

The Heaviside gyromagnetic forces can dominate in the Einstein stationary geodesics (2) in the inner zone of rotating densities, where 
$4r(ln g_o)^\prime_r \ll g^2[1+r(ln [g/g_o])^\prime_r]^2 $
$ \approx {v^2_{+}}/{c^2}, {v^2_{-}}/{c^2} \ll {v^2_{+}}/{c^2} \approx g^2 \ll 1, $  and 
\begin {eqnarray}
\frac {v_{\pm}}{c} = 
\frac {g[1\!+\!r(ln [g/g_o])^\prime_r]}{2}\!
\pm\! \sqrt {  \frac {g^2[1\!+\!r(ln [g/g_o])^\prime_r]^2}{4}\! +\!r(ln g_o)^\prime_r} 
\cr	
\approx	\cases{ 
g[1+r(ln [g/g_o])^\prime_r] + 	\frac {r(ln g_o)^\prime_r}{g[1+r(ln [g/g_o])^\prime_r]} \approx g ,	  \cr 
-	\frac {r(ln g_o)^\prime_r}{g[1+r(ln [g/g_o])^\prime_r]} \approx -\frac {c r(ln g_o)^\prime_r}{ v_{+}}.
}
\end{eqnarray}
In this inner zone of rotating orgabizations even small gyro-potentials, $c^2 g(r)$ $\ll c^2$, in the geodesic solutions (4) can significantly change the Keplerian angular frequencies  $\pm c\sqrt { (\partial_r ln g_o) /r}$ by the Zeeman-type shift $cg(r)/r$ if $g^2(r)\geq 4r\partial_r g_o(r)$. 
Metric gyro-potentials increase almost linearly with the rotation radius within spinning planets and dense regions of disk galaxies, reaching maximum values in their near zone ($r\approx r_{nz}$), where $\partial_r g_{nz}(r) \approx 0 $, $  r \partial_r ln g_o \ll g^2_{nz} \approx const \ll 1$,  $|r \partial_r ln (g_{nz}/g_o)|\ll 1$,  $v_{+} \approx cg_{nz} \ll c, (-v_{-}) \approx c^2 r \partial_r ln g_o /v_{+} \ll v_{+}$.

The exact quantitative distributions of the scalar and gyrometric potentials correspond to the exponentially decreasing densities observed in the disc of the rotating galaxy.  So far GR specialists have been unable to calculate the metric of the rotating galaxy analytically. Qualitatively, however, it can be seen that the vector and scalar potentials could grow radially within the inner zone. Beyond the dense disc, the vector potential slowly falls to zero ($c^2g(r\gg r_{nz})\rightarrow0$) and the scalar potential continues to grow to  saturation ($c^2g_o(r\gg r_{nz})\rightarrow c^2$) in the far zone.  Observable deviations of direct and tetrograde orbites from Kepler\rq{s} laws can be expected from (4) in the inner zone of self-gravitating distributions with differential rotation. These deviations  can be measured not only in galaxies or Saturn\rq{s} rings, but also in the vicinity of rapidly rotating giant planets with circulating moons.

\section {Conclusions} 

Instrumental observations of different speeds on direct and retrograde free paths can be suggested in the near zone of rotating planets, the Kuiper belt and the dense edges of galactic disks. Newtonian dynamics in a one-component gravitational potential and Einstein\rq{s} geodesics in a four-component potential are not identical theories for low-speed satellites and spacecrafts. In contrast to Newtonian gravity, Einstein\rq{s} metric dynamics predicts the Zeeman-like shift of angular frequencies for both direct and retrograde orbits around galactic disks and rotating planets.

The Lorentzian structure of gyroforces in mechanical motion can be tested not only against the predicted difference between direct and retrograde orbits, but also against other phenomena of celestial mechanics. According to Einstein-Grossmann's geodesics, a statistical analysis of seasonal meteor showers should reveal the gravitational manoeuvres of these space bodies to be asymmetric when approaching rotating planets from opposite meridian sides.








\medskip 
 {\bf Funding information}
 
  Author acknowledges support from the Russian Science Foundation (grant No. 23-72-30002).




\end{document}